\documentclass[9pt,twocolumn]{article}
\usepackage[a4paper,margin=1.8cm]{geometry}
\usepackage{multicol}
\usepackage{amsmath,amssymb}
\usepackage{graphicx}
\usepackage{xcolor}
\usepackage{gensymb}
\usepackage{lineno}

\title{Multimodal Speckle-polarization Fiber-optic Sensing for Localized and High-bandwidth Vibration Monitoring}

\author{
Catarina S. Monteiro$^{1,+}$, 
Tomás Lopes$^{1,2,+,*}$, 
Joana Teixeira$^{1,2}$, 
Tiago Ferreira$^{1,2}$, \and
Pedro A. S. Jorge$^{1,2}$, 
Nuno A. Silva$^{1,2,+}$
}
\date{\small
$^1$INESC TEC, Porto, Portugal \\
$^2$Departamento de Física e Astronomia, Universidade do Porto, Portugal \\
$^+$These authors contributed equally \\
$^*$tomas.j.lopes@inesctec.pt}

\begin{document}
\maketitle

\begin{abstract}
    High-bandwidth and multi-point acoustic and vibration sensing is a critical asset for real-time condition monitoring, maintenance, and surveillance applications. In the case of large scales and harsh environments, optical fiber distributed sensing has emerged as a compelling alternative to electronic transducers, featuring lower installation and maintenance costs, along with compact footprints and enhanced robustness.  Yet, current distributed fiber-optic sensing solutions are typically costly and face a resolution-bandwidth tradeoff. In this work, we present an alternative fiber-optic vibration sensing strategy that harnesses a multimodal architecture combining speckle and polarization interrogation. The experimental results demonstrate the concept by achieving speckle-based signal source localization with centimeter-range spatial resolution, while obtaining a high-fidelity waveform reconstruction over at least 100 Hz–40 kHz bandwidth via the polarimetric sensing part. Overall, the work establishes a general and promising blueprint to harness multimodality in fiber sensing and break single-modality constraints.
\end{abstract}

%%%%%%%%%%%%%%%%%%%%%%%%%%%%%%%%%%%%%%%
%%%%%%%%%%%%%%%%%%%%%%%%%%%%%%%%%%%%%%%

%%%%%%%%%%%%%%%%%%%%%%%%%%%%%%%%%%%%%%%
%%%%%%%%%%%%%%%%%%%%%%%%%%%%%%%%%%%%%%%

\section{Introduction}
As industrial systems and smart cities move toward autonomous operation, scalable acquisition of broadband acoustic and vibration data becomes essential for real-time situational awareness and model-based diagnostics. For example, the preservation of critical infrastructure such as railways, power grids, oil and gas pipelines requires continuous monitoring of the whole infrastructure to detect possible leaks, unauthorized intrusions, or structural degradation. Real-time, continuous monitoring is essential for lifecycle maintenance, aiming to prevent unnecessary disruptions to operations. Conventional solutions rely on discrete transducers such as arrays of geophones, accelerometers, and hydrophones that, even though they offer high sensitivity, have high installation and maintenance costs, are typically bulky and heavy, and can only provide point-wise coverage \cite{Dai2015, Liu2023}. Recently, the usage of unmanned aerial vehicles has also been explored for infrastructure health monitoring, proven invaluable for inspecting hard-to-reach sections of critical infrastructure \cite{Aela2024}. However, their use remains confined to periodic surveys rather than true continuous monitoring. Constraints on flight duration and battery life limit the working time, while payload restrictions and data-processing limits the number of onboard sensors \cite{Mohsan2023}. This means that, despite combining multiple sensing modalities with drone mobility, a single UAV cannot yet deliver real-time, multi-point monitoring over extended periods. In this scenario, optical fiber sensors emerge as a compelling alternative in the form of a compact and lightweight solution, with high resilience to harsh environments, and inherent multiplexing and multi-point capabilities.

Indeed, for multi-point vibration and acoustic monitoring, there is already a well-established market with a broad spectrum of fiber-optic technologies. The most common are point and quasi-distributed sensors based on fiber Bragg gratings, Fabry–Perot cavities, or interferometric probes deliver high-SNR readout at selected locations via wavelength or time-division multiplexing\cite{lu2019distributed, li2021fbg, shao2025artificial, rahman2024review, ma2023review}. However, these depend on special fibers or components, meaning an additional per-sensor cost, and the need for multiplexing interrogation imposes a trade-off between bandwidth and spatial resolution. On the other hand, operating in standard telecom fibers, distributed fiber sensing has gathered significant attention within the scientific and engineering academic communities. Today, coherent Rayleigh ($\phi$-OTDR or DAS) is already widely used for dynamic strain and acoustic sensing \cite{he2021optical, guan2020high}, while Brillouin/Raman methods excel at quasi-static strain and temperature \cite{layosh2025forward, antman2016optomechanical, wang2017few}. Yet, these distributed schemes again face trade-offs among range, spatial resolution, and temporal bandwidth. Besides, these also often demand complex and costly interrogators, with a non-negligible computational footprint that hinders real-time and sensing edge applications.

In this context, an interesting way to bypass the pressing resolution-bandwidth tradeoff may be to combine sensing principles into a multimodal solution. Among diverse alternatives that work in standard optical fibers, speckle-based sensing and state-of-polarization (SoP) interrogation emerge as particularly promising due to their complementary strengths. On one hand, in speckle-based fiber sensing, coherent light is propagated through a multimode fiber (MMF), which supports the guidance of multiple spatial modes. The interference between these modes produces a complex, granular intensity distribution known as a speckle, which is known to be highly sensitive to external perturbations such as strain, vibration, or temperature changes \cite{Fujiwara2019, Murray2019}. Moreover, as different points on the fiber affect differently the speckle pattern, localized disturbances manifest as distinct, time-evolving pattern shifts that can be tracked with high accuracy using imaging sensors \cite{Shimadera2022, Gao2023}. However, the temporal resolution of speckle sensing is fundamentally limited by the frame rate of the camera, typically restricting signal reconstruction bandwidths to a few kilohertz \cite{Murray2019}. On the other hand, SoP interrogation allows broadband reconstruction of waveforms up to the MHz range by exploiting birefringence changes caused by external perturbations \cite{Charlton2017, Costa2023, Carver2024}. Changes in local birefringence cause rotation and ellipticity changes in the polarization state of the transmitted light, which can be fully described by the Stokes vector. Yet, the measured polarization reflects the cumulative effect along the entire length of the fiber, meaning that the technique alone cannot pinpoint the exact location of the perturbation.

Nevertheless, one may conjecture that the complementary advantages between the two techniques, i.e., the high spatial resolution provided by the speckle sensing and the wideband waveform reconstruction provided by the SoP, may be combined into a multimodal approach that can overcome the individual limitations of each technique. This conceptual idea, which strongly aligns with the emerging concept of sensor fusion, establishes the starting point for the present manuscript, where we introduce a multimodal fiber optic sensor that combines speckle dynamics with SoP interrogation into a unified platform. Establishing a theoretical framework for each modality, we first develop a mathematical methodology to support real-time localization via speckle analysis, whereas SoP is utilized for high temporal resolution reconstruction of the vibration signal. Experimentally, we show that speckle-pattern shifts captured by a standard high-speed camera enable spatial localization of non-overlapping temporal perturbations with a spatial resolution of approximately 3\,cm at least, while a compact high-speed polarimeter measures the time-varying Stokes vector to reconstruct the acoustic waveform with a bandwidth exceeding 40~kHz. By decoupling spatial sensitivity from temporal resolution, this architecture overcomes the trade-offs that constrain single-modality systems, paving the road not only for direct applications requiring both high fidelity and high spatial granularity (e.g., pipelines or rail sections), but also for new sensing strategies following a similar multimodal blueprint.

%%%%%%%%%%%%%%%%%%%%%%%%%%%%%%%%%%%%%%%
%%%%%%%%%%%%%%%%%%%%%%%%%%%%%%%%%%%%%%%

%%%%%%%%%%%%%%%%%%%%%%%%%%%%%%%%%%%%%%%
%%%%%%%%%%%%%%%%%%%%%%%%%%%%%%%%%%%%%%%

\section{Modeling Speckle Dynamics and Polarization in Multimodal Fiber Sensing}
\label{sec:multimodal}

In this section, we develop a mathematical framework for multimodal fiber sensing that treats two complementary observables as two linearly perturbation-sensitive channels acted on by the same physical disturbance. By modeling the effect on the multimode speckle intensity field recorded by an imaging sensor and on the state of polarization (SoP) measured by a polarimeter, we make explicit how each observable encodes different aspects of the perturbation and encloses distinct limitations, showing also how the fusion of the two readouts can overcome them.

%%%%%%%%%%%%%%%%%%%%%%%%%%%%%%%%%%%%%%%
%%%%%%%%%%%%%%%%%%%%%%%%%%%%%%%%%%%%%%%

%%%%%%%%%%%%%%%%%%%%%%%%%%%%%%%%%%%%%%%
%%%%%%%%%%%%%%%%%%%%%%%%%%%%%%%%%%%%%%%

\subsection{Localization of Perturbations via Speckle Dynamics}
\label{sec:speckle_monitoring}

Multimode fibers (MMFs) support multiple propagation modes, each with distinct propagation constants, spatial modes, and phase velocities \cite{Goodman2007}. At the output of the fiber, the coherent superposition of these modes, combined with mode mixing and modal dispersion, results in the generation of a complex interference pattern known as speckle. Although the speckle pattern appears to be random, it has a deterministic behavior for fixed conditions and is highly sensitive to environmental conditions that alter relative modal phases or inter-modal coupling, such as temperature variations or mechanical vibrations \cite{Rahmani2018}. 

To model these dynamics, we first notice that the spatial intensity of the speckle can be determined by the decomposition of the intensity of the superposition of the $N$ guided modes as
\begin{equation}
    I(r,t) = |E(r,t)|^2 = \left| \sum_j^N a_j(t)\psi_j(\mathbf{r}) exp(i\phi_j(t))\right|^2
    \label{spatial_intensity}
\end{equation}
where $E$ represents a time-dependent monochromatic field propagating along the fiber, $a_j$ and $\phi_j$ are the time-dependent amplitude and phase of the mode $j$, and $\psi_j$ represents the spatial field distribution of each mode. Under weak perturbations, we can obtain a linear contribution
\begin{equation}
   \Delta I \approx 2 \sum_{j,k} a_j a_k \cos(\phi_j - \phi_k)\Delta\phi_{j,k}
   \label{speckle_approx}
\end{equation}
where $\Delta\phi_{j,k}$ represents the inter-modal coupling changes induced by the induced perturbation, thus encoding the information about environmental changes into the high-dimensional speckle response \cite{Gutierrez-Cuevas2024}. Techniques such as analog or digital phase conjugation \cite{Dunning1982, Papadopoulos2012}, digital iterative methods \cite{DiLeonardo2011, Cizmar2012}, and, more recently, the use of deep learning \cite{Rahmani2018, Zhu2021} have been employed to describe light propagation along complex media. 

Yet, to avoid the complex problem of modeling both the spatial modes and the coupling induced by the perturbation, a simpler way to approach these complex dynamics is to utilize the transmission matrix formalism to directly map the output spatial intensity from a known input spatial intensity pattern using an experimentally determined transmission matrix $\mathbf{M}$. Under this formalism, the effect of a localized perturbation $\delta$ at point $x$ on the output optical field can be expressed as
\begin{equation}
    \mathbf{E_{out}}(x; \delta) = \mathbf{\bar M_2 \bar D}(x;\delta) \mathbf{\bar M_1} E_{in}
    \label{output_field}
\end{equation}
where $\mathbf{\bar M_{1,2}}$ are respectively the unperturbed transmission matrices describing light propagation through the MMF before and after the disturbance at position $x$, and $\mathbf{\bar D }(x;\delta)$ captures the change in the transmission matrix induced by a small perturbation of intensity $\delta$. In the linear regime, i.e. small $\delta$, the matrix $\mathbf{\bar D}(x;\delta)$ can be linearized as $\mathbf{\bar D(x) \approx \bar D_0} + (\partial_\delta \mathbf{\bar D})\delta$ where $\mathbf{\bar D_0}$ represents the transmission matrix with no perturbation. Substituting into equation \ref{output_field}, yields $\boldsymbol{E}_{out} \approx \boldsymbol{E}_o(x) + \delta \boldsymbol{E}_1(x)$, with $\boldsymbol{E}_0(x) = \mathbf{\bar M_2 \bar D_0(x) \bar M_1 \boldsymbol{E}_{in}}$ and $\boldsymbol{E}_1(x) = \bar{ \boldsymbol{M}_2} (\partial_\delta \bar{\boldsymbol{D}}(x))\bar{ \boldsymbol{M}_1} \boldsymbol{E}_{in}$. Expressed as intensity, this gives
\begin{equation}
    \mathbf{I}(x;\delta) \approx \boldsymbol{I}_0(x) + (\partial_\delta \boldsymbol{I}(x))\delta + \mathcal{O}(\delta^2)
\end{equation}
where $\mathbf{I}$ can now be associated with a vector $\mathbf{I}(x) \in \mathbb{R}^m $ with $m$ representing the number of macropixels in the detector, and $\boldsymbol{I}_0$ is the intensity recorded at the camera without perturbation. The partial derivative term can be associated with the formal definition
\begin{equation}
    \partial_{\delta} \mathbf{I}(x) = \lim_{\delta \rightarrow0} \frac{\mathbf{I}(x,\delta) - \mathbf{I}_0}{\delta}.
\end{equation}

For sufficiently small $\delta$, the speckle pattern thus evolves linearly with the perturbation, enabling straightforward extraction of $\delta$ from measured intensity changes. This can be generalized for a set $\{\delta_i \}$ of $N$ deformations at different $x_i$ points as
\begin{equation}
    \boldsymbol{I}(\{x_i,\delta_i\}) \approx I_0 + \sum_{i=1}^N (\partial_{\delta_i} \mathbf{I}) \delta_i
\end{equation}
with $\partial_{\delta_i}\mathbf{I}$ being a partial derivative given by 
\begin{equation}
    \partial_{\delta_i} \mathbf{I} = \lim_{\delta_i \rightarrow0} \frac{\mathbf{I}(0,~...~, \delta_i, ~...~,0) - \mathbf{I}_0}{\delta_i}.
\end{equation}
Each deformation signal $\delta_i$ (which can have a temporal variation) can then be reconstructed from local intensity variations as
\begin{equation}
\begin{aligned}
    \begin{bmatrix}
        \delta_1\\
        \vdots\\
        \delta_N
    \end{bmatrix} &\approx 
    \bigl[\boldsymbol{I}(\delta_1, ..., \delta_N) - \boldsymbol{I}_0\bigr] \cdot \frac{ (\partial_\delta \boldsymbol{I})^\dagger }
    {\bigl\lVert (\partial_\delta \boldsymbol{I})^\dagger \bigr\rVert^2}.
\end{aligned}
\label{eq:def}
\end{equation}
where the $\dagger$ symbol denotes the Moore-Penrose pseudo-inverse and $    \partial_\delta I \equiv [\partial_{\delta_1} \mathbf{I} ~...~\partial_{\delta_N}\mathbf{I}] ^{T}$.

From this mathematical framework, one can directly conclude that localized vibration sensing is possible and viable due to the large dimensionality of the output speckle, with the maximum number of separable spatial points being associated with the rank of the Jacobian matrix $\partial_\delta \mathbf{I}$, which theoretically correspond to the minimum of the number of spatial modes or the number of detector macropixels. Yet, in real-world conditions, reconstructing the temporal variation of detected deformation using speckle-based sensing is highly limited due to constraints on the interrogation side. Indeed, the limited frame rate and dynamic range of standard cameras, along with their latency, limited memory bandwidth, and necessary computational load, make the approach unsuitable for monitoring applications that operate at kilohertz rates.

%%%%%%%%%%%%%%%%%%%%%%%%%%%%%%%%%%%%%%%
%%%%%%%%%%%%%%%%%%%%%%%%%%%%%%%%%%%%%%%

%%%%%%%%%%%%%%%%%%%%%%%%%%%%%%%%%%%%%%%
%%%%%%%%%%%%%%%%%%%%%%%%%%%%%%%%%%%%%%%

\subsection{High-Bandwidth monitoring via Polarization dynamics in optical fibers}
\label{sec:polarization_monitoring}

Polarimetric interrogation is a well-established fiber-optic sensing technique noted for very high sensitivity and inherently large temporal bandwidth. In conceptual terms, as coherent light propagates through a standard single-mode optical fiber, intrinsic defects and extrinsic perturbations such as mechanical strain, temperature fluctuations, or acoustic wave-induced vibrations, modify the birefringence and polarization-mode coupling of the fiber. These perturbations modulate the differential phase between orthogonal polarization modes and induce polarization-mode coupling, driving the state of polarization (SoP) of the transmitted field to evolve on the Poincaré sphere.

To put on a mathematical framework, the complete polarization state can be described by the Stokes vector $\mathbf{S_{out}} = [S_0, S_1, S_2, S_3]^T$, where $S_0$ is the total intensity, $S_1$ and $S_2$ represent linear polarization components at $0\degree/90\degree$ and $\pm45\degree$, respectively, and $S_3$ corresponds to the circular polarization component. When a perturbation $\delta$ is applied to the fiber, it induces a corresponding variation in the output Stokes vector from its unperturbed state $\mathbf{S_0}$. Provided the perturbation is sufficiently small, the output polarization state $\mathbf{S}_{out}$ can be linearized using a first-order Taylor expansion:
\begin{equation}
    \mathbf{S}(t) \approx \mathbf{S_0} + (\partial_\delta \mathbf{S})\delta(t)
\end{equation}
where $\partial_\delta S$ denotes the sensitivity matrix of the Stokes vector with respect to the applied perturbation, which is determined by applying known perturbations to the fiber and recording the corresponding changes in the Stokes components. This relation establishes the direct estimation of the perturbation time series $\delta(t)$ from the measured polarization changes as
\begin{equation}
    \delta(t) \approx (\mathbf{S}(t) - \mathbf{S}_0) . \frac{(\partial_\delta \mathbf{S})^\dagger}{|| (\partial_\delta \mathbf{S})^\dagger ||^2}
\label{polarization-perturbation}
\end{equation}
where $\dagger$ denotes the Moore-Penrose pseudo-inverse of the sensitivity matrix. 

As can be seen, the formalism closely follows the one obtained in the previous section for the speckle case. Yet, it shall be noted that the output SoP lies on a sphere (neglecting variations of $S_0$) rather than on a high-dimensional macropixel space as in the previous section. This means the effective dimensionality of the SoP output space and thus the rank of the Jacobian matrix $\partial_\delta \mathbf{S}$ is just 2. Thereby, unlike the speckle case, polarimetric sensing does not support scalable spatial resolution along the fiber. Also, considering non-overlapping time signals, there is no formal need to pre-calibrate the Jacobian $\partial_\delta \mathbf{S}$, as one may substitute the vector on the right-hand side of equation \ref{polarization-perturbation} by a random vector and still likely get a weighted integration of all the effects along the fiber in just one signal.

%%%%%%%%%%%%%%%%%%%%%%%%%%%%%%%%%%%%%%%
%%%%%%%%%%%%%%%%%%%%%%%%%%%%%%%%%%%%%%%

\subsection{Harnessing Multimodal Speckle and Polarization Sensing}

Having established the individual principles and capabilities of speckle dynamics and SoP interrogation, we now combine these modalities into a unified sensing architecture aiming at overcoming the limitations of using either technique alone. 

As we saw, speckle dynamics is able to both retrieve and locate signals along the fiber, but features low bandwidth. In practice, even though it is still capable of locating signals at high frequencies if in the linear regime (even entering the ultrasonic range), signal reconstruction is still limited by the Nyquist limit to a few kHz in a best-case scenario. Conversely, SoP interrogation is not capable of distinguishing or spatially separating multiple events: the lack of dimensionality of the output space means that it aggregates all perturbations into a single polarization measurement. Yet, through SoP interrogation, we can capture signals at very high acquisition rates, with signal reconstruction capabilities at frequencies extending into the MHz regime. 

Aiming to capitalize on the strength of each modality, we propose in this work to combine the two techniques in a multimodal sensing solution that allows both localization and high-bandwidth vibration sensing. For this purpose, we utilize a two-stage hyphenated sensor fusion architecture, leveraging continuous SoP interrogation to monitor existing vibration signals, and speckle dynamics to locate perturbations on the fiber with a spatial resolution down to the centimeter scale. 

The first processing stage concerns the SoP interrogation part. As we saw in the previous section, the optimal conditions to reconstruct the signal require a suitable calibration vector $\partial_\delta \boldsymbol{S}$ determined for each of the sensing points. Yet, as we also discussed in section \ref{sec:multimodal}\ref{sec:polarization_monitoring}, the low dimensionality of the SoP and lack of localization capabilities mean that a non-optimal interrogation vector may be sufficient. Therefore, experimentally we can approximate $\partial_\delta \boldsymbol{S}$ by a finite difference derivative
\begin{equation}
    \partial_\delta \boldsymbol{S}_{exp} \approx \frac{\boldsymbol{S}(\delta_1,...,\delta_N)-\boldsymbol{S}_0}{\delta}
    \label{pol_calibration}
\end{equation}
computed with every location vibrating with the same signal $\delta_1$. With this vector, we are then able to reconstruct the total vibration signal $\delta(t)_{total,rec}$ using equation \ref{polarization-perturbation}.

In the second stage, we proceed to the localization using the speckle part. For this, we must identify optimal interrogation modes that effectively separate signals from different spatial locations, enabling precise localization of fiber deformations. As described in section \ref{sec:multimodal}\ref{sec:speckle_monitoring}, each optimal interrogation mode corresponds to a row of the sensitivity / Jacobian matrix $\partial_\delta\boldsymbol{I}$, which can experimentally be determined by approximating the derivative of the speckle pattern for a given calibration set of applied localized deformations via finite differences as
\begin{equation}
    \partial_{\delta_i} \boldsymbol{I}_{exp} \approx \frac{\boldsymbol{I}(0,...,\delta_i,...,0)-\boldsymbol{I}_0}{\delta_i}.
\end{equation}
Having the sensitivity matrix, one may reconstruct the temporal dynamics of the measured deformation $\delta(t)_{i,rec}$ for each $i$-th channel, and localize the signal according to a given score. For the purpose of this work, we computed the score via the instantaneous power of the signal, defined from the spectrogram as

\begin{equation}
    P_i(t) = \int_{-\infty}^{+\infty}|X_i(t, f)|^2 df = \int_{-\infty}^{+\infty}S_i(t, f)df,
\end{equation}

where $X_i(t, f)$ denotes the short-time Fourier transform (STFT) of signal $i$, with the spectrogram $S_i(t, f) = |X_i(t, f)|^2$. By Parseval's theorem applied to the STFT, the previous signal equals, up to a constant, the windowed time-domain energy.

%%%%%%%%%%%%%%%%%%%%%%%%%%%%%%%%%%%%%%%
%%%%%%%%%%%%%%%%%%%%%%%%%%%%%%%%%%%%%%%

\section{Results}

\begin{figure}
    \centering
    \includegraphics[width=0.9\linewidth]{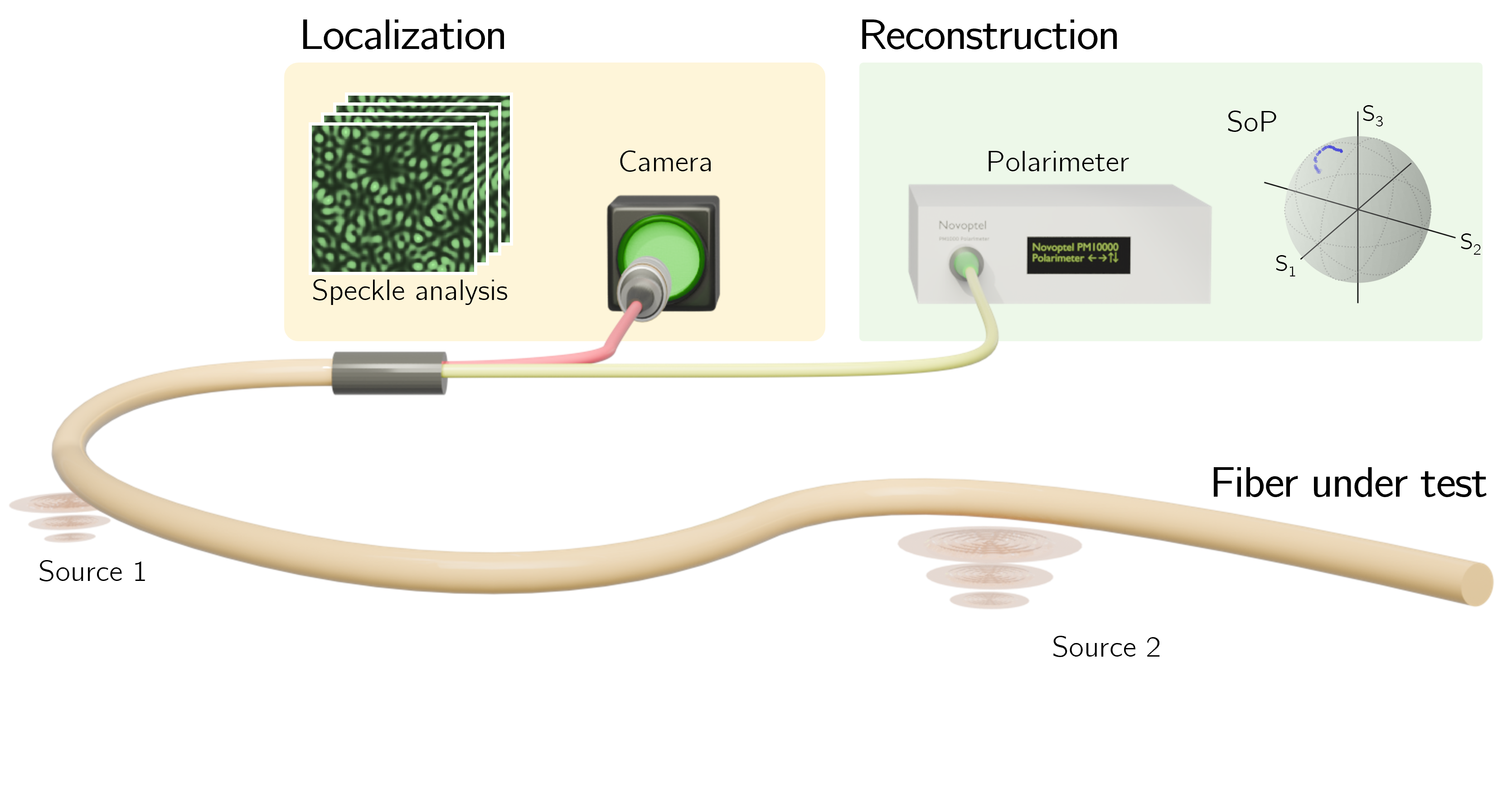}
    \caption{Multimode optical fiber interrogation system composed by two parallel paths: the speckle-localization module, composed by a 532\,nm laser, a multimode fiber, and a high-speed camera that images evolving speckle patterns to localize the events and the reconstruction module composed by a 1500\,nm laser, a single mode fiber and a polarimeter to interrogate the full-Stokes parameters and reconstruct the waveforms up to 40~kHz.}
    \label{fig:setup}
\end{figure}

To implement and demonstrate the capabilities of the described framework, we deployed the multimodal experimental setup shown in Figure \ref{fig:setup}. The system comprises two parallel optical interrogation arms that follow the same physical path through four piezoelectric membrane actuators.

The first arm utilizes speckle pattern analysis, employing a 532\,nm laser (CNI, MSL-DS-532) coupled into a 4-meter-long step-index multimode optical fiber (50 $\mu \text{m}$ core diameter, 125\,$\mu \text{m}$ cladding diameter, 0.2\,NA). The transmitted light is captured by a CMOS camera (Ximea, MQ013MG-ON) for speckle pattern interrogation. The capture was limited to a region of interest of $64 \times 64$ pixels, allowing us to achieve a framerate of 4000\,FPS, which, per Nyquist's theorem, translates into a maximum frequency reconstruction up to 2\,kHz.

The second arm employs polarization analysis, featuring a 1550\,nm laser source (Santec, TSL-210) coupled into a 4-meter-long single-mode optical fiber (9.2\,$\mu \text{m}$ core diameter, 125\,$\mu \text{m}$ cladding diameter, 0.14\,NA). The polarization state of the transmitted light is analyzed using a Novoptel PM1000 polarization interrogation unit with a maximum sampling rate of 100\,MHz.

Both optical fibers are routed along identical paths, passing through four piezoelectric membrane speakers (TDK, PHUA3015-049B-00-000), separated by different distances (3\,cm to 1.2\,m), that serve as controlled acoustic vibration actuators. The control is achieved via a National Instruments USB-6259 system, enabling the generation of acoustic disturbances for sensing validation.

As described in the previous section, the operational mode of this sensor requires establishing appropriate calibration procedures for both interrogation methods. To achieve these objectives, we implemented a two-step calibration process. For the speckle interrogation arm, individual 500\,Hz sinusoidal signals were sequentially applied to each of the four piezoelectric actuators. By synchronizing the input and output signals and taking the frame at the maximum of the input signal, the approach enables the construction of a sensitivity matrix row-by-row by approximating the derivatives of the speckle pattern with respect to localized deformations via finite differences. Under the first-order approximation, computing the pseudo-inverse of this sensitivity matrix allows reconstruction of the deformation applied at each actuator location with minimal crosstalk between channels according to equation \ref{eq:def}. On a second step, for the SoP arm, we drove the piezoelectric actuators with a simultaneous signal of 500\,Hz and computed the finite difference approximation as in equation \ref{pol_calibration}.

%%%%%%%%%%%%%%%%%%%%%%%%%%%%%%%%%%%%%%%
%%%%%%%%%%%%%%%%%%%%%%%%%%%%%%%%%%%%%%%

\subsection{Frequency Characterization}
Following calibration of both interrogation modalities, we proceed to characterize the frequency response capabilities of each sensing arm. For this, a single piezoelectric actuator was driven with a linear frequency sweep from 100\,Hz to 40\,kHz over a 20-second duration. The chirp signal reconstruction was performed using the established calibration procedures: equation \ref{polarization-perturbation} for the polarization arm and the corresponding sensitivity matrix inversion (equation \ref{eq:def}) for the speckle interrogation arm.

\begin{figure}
    \centering
    \includegraphics[width=\linewidth]{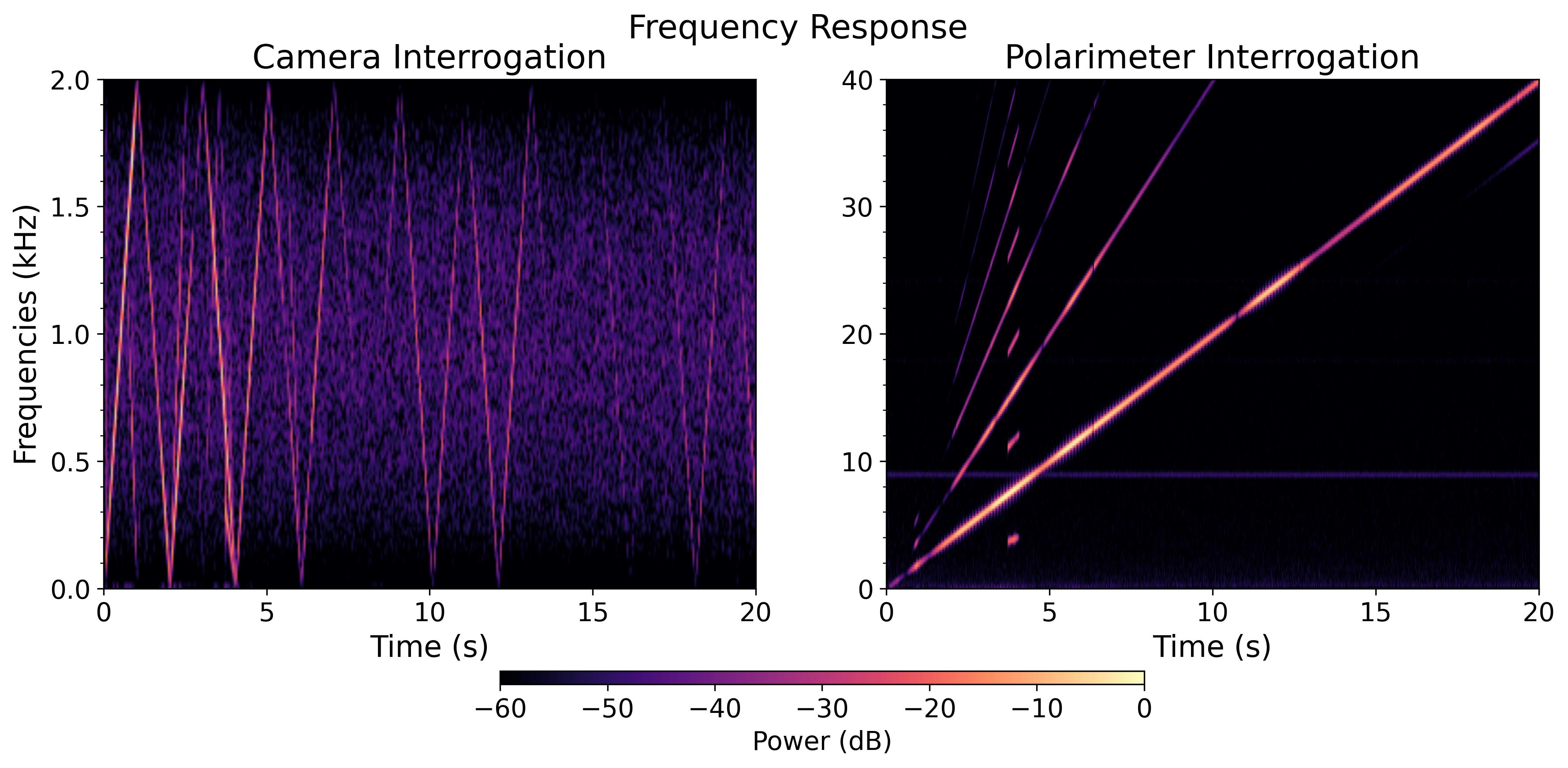}
    \caption{Frequency response characterization of the dual-modality sensing framework. The image on the left shows the speckle-based interrogation response (4000\,FPS sampling), exhibiting aliasing beyond the 2000\,Hz Nyquist frequency. The image on the right presents the polarization-based interrogation response (200\,kHz acquisition rate), demonstrating linear frequency reconstruction across the complete 100\,Hz to 40\,kHz range.}
    \label{fig:frequency_response}
\end{figure}

Figure \ref{fig:frequency_response} presents the reconstructed frequency responses obtained for each modality, normalized to the maximum power registered in each spectrogram. Starting with the speckle-based interrogation, it is straightforward to see that the 4000\,FPS sampling rate imposes a Nyquist frequency limit of 2000\,Hz, resulting in aliasing artifacts for frequencies beyond this threshold. Consequently, the reconstructed signal does not faithfully reproduce the linear frequency sweep for the complete frequency range, as expected. Yet, for the intended operational paradigm of the speckle modality, i.e., localization of the signal, the results suggest that the method may still serve as a spatial localization tool for fiber deformations even beyond the Nyquist frequency. Indeed, the presence of spectral aliasing and signal reflections at the Nyquist boundary can still be utilized as an indicator of external perturbations affecting the speckle dynamics, thereby enabling deformation detection without requiring accurate frequency reconstruction.

Advancing to the polarization interrogation arm, which was operated at an acquisition rate of 200\,kHz, the results demonstrate superior frequency response capabilities across the entire tested range. Indeed, the reconstructed signal shows a good linearity throughout the 40\,kHz sweep with a high signal-to-noise ratio, demonstrating that this modality is, in fact, well-suited for high-fidelity vibration reconstruction over the frequency band of interest. Additionally, comparing the response of the two modalities reveals a reduced signal-to-noise ratio in the camera-based interrogation system, with certain frequency bands exhibiting signal levels approaching the noise floor. This suggests that the polarization-based interrogation exhibits higher sensitivity to mechanical vibrations, which can be attributed to lower coupling efficiency between mechanical deformations and speckle pattern modulations under the given experimental configuration. We shall stress, however, that this constraint can be mitigated through optimization of the fiber-perturbation coupling mechanism within the operational frequency range of interest, thereby enhancing the mechanical-to-optical coupling efficiency for this modality.

\subsection{Localization of Perturbations}
Having demonstrated the capabilities of this framework to detect perturbations across a wide frequency range, we now proceed to test the ability of the system to spatially localize deformations and correctly identify their position along the fiber. This was accomplished by driving each of the four piezoelectric actuators with distinct frequencies (8.3\,kHz, 1.3\,kHz, 6.7\,kHz, and 35\,kHz) for a duration of one second, creating known deformations at specific fiber locations.

\begin{figure}
    \centering
    \includegraphics[width=\linewidth]{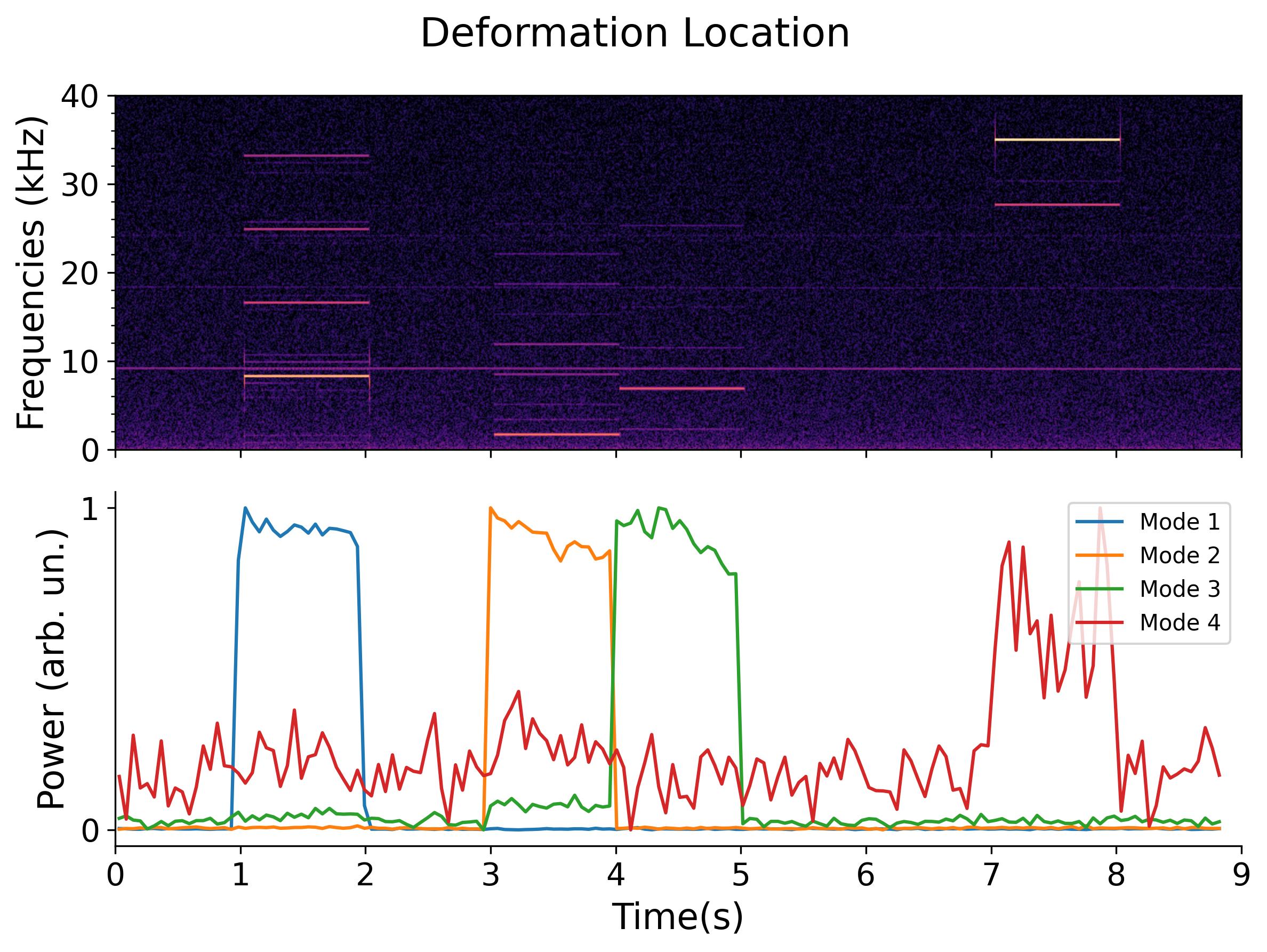}
    \caption{Localization performance of the multimodal sensing framework. Each actuator was driven with a distinct frequency for one second. The image on top shows the global perturbations happening across the entire fiber. The image at the bottom displays the normalized power spectral density, which shows the ability of the system to correctly identify the active deformation location.}
    \label{fig:localize_deformations}
\end{figure}

Signal reconstruction was performed using the same methods established during calibration. For the polarization modality, we applied equation \ref{polarization-perturbation} to reconstruct the aggregate deformation signal. For the speckle interrogation, we used Welch's method to estimate the power spectral density at each deformation mode, with results normalized to the 0-1 range to enable direct comparison between modes.
Figure \ref{fig:localize_deformations} shows the localization results for each actuator position. The normalized power distribution indicates which actuator was active during each measurement interval, demonstrating the capability of this framework to correctly identify deformation locations with minimal crosstalk between channels. Here, it is easy to identify that the signal played at Mode 4 (35\,kHz) was near the noise level, yet in comparison to the other modes, it is still possible to correctly identify the location of the deformation.

\begin{figure}
    \centering
    \includegraphics[width=0.65\linewidth]{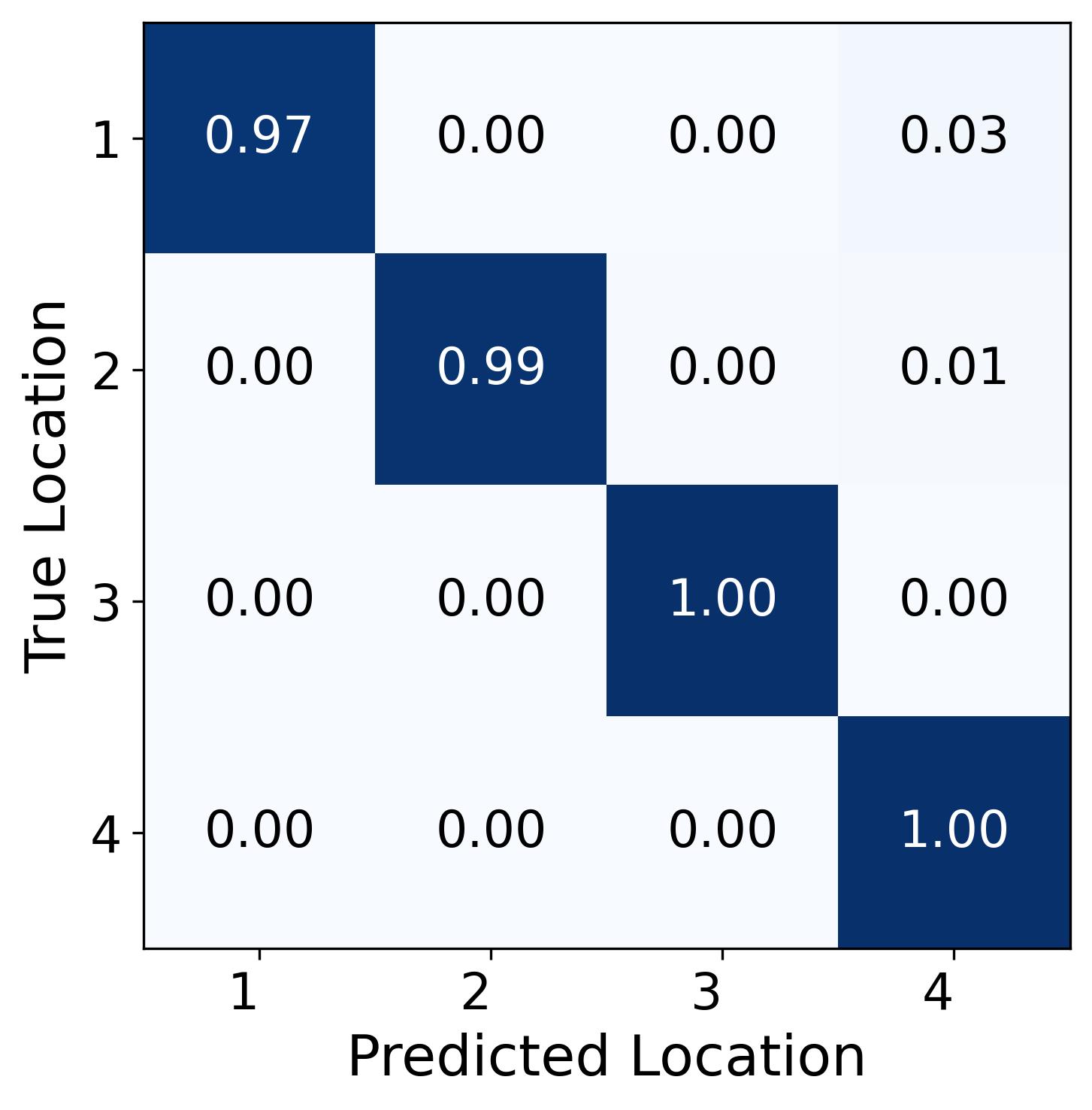}
    \caption{Confusion matrix demonstrating localization accuracy across 1000 perturbation events distributed over four actuator positions.}
    \label{fig:localization_confusion_matrix}
\end{figure}

Finally, to better quantify the localization accuracy and assess system stability over extended operation periods, we constructed a confusion matrix based on 1000 perturbation events distributed across the four actuator positions. The test signals comprised 100 randomly selected one-second excerpts from six royalty-free music clips, with the complete data acquisition spanning approximately 22 minutes. Figure \ref{fig:localization_confusion_matrix} presents the classification results, showing that the system correctly predicts deformation positions with near 100\% accuracy, serving as a good demonstration of the stability and robustness of the sensing framework.

\section{Concluding Remarks}
In this work, we presented a multimodal sensing framework that combines speckle dynamics with state-of-polarization (SoP) interrogation to overcome some of the limitations of existing fiber-optic multi-point vibration sensing approaches. By integrating these two complementary techniques, we achieve both centimeter-scale localization and high-fidelity deformation reconstruction across frequencies from 100 Hz to 40 kHz, addressing the traditional trade-off between spatial resolution and temporal bandwidth that constrains single-modality systems. 

Using a proof-of-concept implementation, the experimental results validate the effectiveness of each sensing modality within the combined framework. The speckle interrogation arm, operating at 4000 FPS through a multimode fiber, successfully localizes perturbations along the fiber with minimal crosstalk between different positions. Using the Moore-Penrose pseudo-inverse reconstruction approach, the system correctly identifies deformation locations with near 100\% accuracy across 1000 test events over extended measurement periods. 
Yet, the frequency characterization reveals that the operational limitation of the speckle-based reconstruction shows aliasing beyond the 2000 Hz Nyquist limit, confirming its role as a localization tool rather than a high-fidelity reconstruction method. 

On the other hand, the polarization interrogation arm maintains linear frequency response throughout the complete 40 kHz range, demonstrating superior sensitivity compared to camera-based speckle analysis and enabling precise vibration reconstruction at high frequencies.

Putting into a broader perspective, the demonstrated multimodal framework establishes a new and scalable multi-point fiber-optic sensing solution, which paves the way for future research into two complementary directions. First, the sensing platform introduced here can be translated to real-world implementations and further investigated in terms of spatial resolution, report strain sensitivity, dynamic range, and long‑term drift. A demonstration on tens to hundreds of meters in relevant environments (e.g., pipelines or rail) would clarify its scalability towards distributed sensing and address practical calibration issues. On a second direction, more on a conceptual level, the work may also inspire new multimodal architectures in the field of fiber-optic sensing, harnessing complementary strengths to bypass current obstacles and limits of existing solutions.

\section{Funding} This work is co-financed by Component 5 - Capitalization and Business Innovation, integrated in the Resilience Dimension of the Recovery and Resilience Plan within the scope of the Recovery and Resilience Mechanism (MRR) of the European Union (EU), framed in the Next Generation EU, for the period 2021 - 2026, within project HfPT, with
reference 41, and by national funds through FCT – Fundação para a Ciência e a Tecnologia, I.P., under the support UID/50014/2023 (https://doi.org/10.54499/UID/50014/2023). Tomás Lopes and Joana Teixeira acknowledge the support of the Foundation for Science and Technology (FCT), Portugal, through Grants 2024.01830.BD and 2024.00426.BD, respectively. Nuno A. Silva acknowledges the support of FCT under the grant 2022.08078.CEECIND/CP1740/CT0002 (https://doi.org/10.54499/2022.08078.CEECIND/CP1740/CT0002).

\section{Acknowledgments} This work is co-financed by Component 5 - Capitalization and Business Innovation, integrated in the Resilience Dimension of the Recovery and Resilience Plan within the scope of the Recovery and Resilience Mechanism (MRR) of the European Union (EU), framed in the Next Generation EU, for the period 2021 - 2026, within project HfPT, with
reference 41, and by national funds through FCT – Fundação para a Ciência e a Tecnologia, I.P., under the support UID/50014/2023 (https://doi.org/10.54499/UID/50014/2023). Tomás Lopes and Joana Teixeira acknowledge the support of the Foundation for Science and Technology (FCT), Portugal, through Grants 2024.01830.BD and 2024.00426.BD, respectively. Nuno A. Silva acknowledges the support of FCT under the grant 2022.08078.CEECIND/CP1740/CT0002 (https://doi.org/10.54499/2022.08078.CEECIND/CP1740/CT0002).

\section{Disclosures} The authors declare no conflicts of interest.

\section{Data Availability Statement} Data underlying the results presented in this paper are not publicly available at this time but may be obtained from the authors upon reasonable request.

\section{Supplemental document}
A supplemental document must be called out in the back matter so that a link can be included. For example, “See Supplement 1 for supporting content.” Note that the Supplemental Document must also have a callout in the body of the paper.

% Bibliography
\bibliographystyle{unsrt} % or abbrv, or IEEEtran, etc.
\bibliography{sample}

\end{document}